\documentclass[aps,twocolumn,prx,showpacs,superscriptaddress,10pt,longbibliography]{revtex4-2}
\pdfoutput=1  
\usepackage{bm}
\usepackage{graphicx}
\usepackage{amsmath}
\usepackage{amssymb}
\usepackage{color}
\usepackage{braket}
\usepackage{wasysym}
\usepackage{enumerate}
\usepackage[normalem]{ulem}
\usepackage{hyperref}
\usepackage{ifthen}
\usepackage{xspace}
\usepackage[caption=false]{subfig}
\usepackage{physics}
\usepackage{braket}

\newcommand{\kB}{k_\text{B}}

\newcommand{\beq}{\begin{equation}}
\newcommand{\eeq}{\end{equation}}

\begin{document}

\title{Asymmetries of thermal processes in open quantum systems}

\author{\'Alvaro Tejero}\email[Corresponding author: ]{atejero@onsager.ugr.es}\affiliation{Electromagnetism and Condensed Matter Department, Universidad de Granada, 18071 Granada, Spain\looseness=-1}%
\affiliation{Instituto Carlos I de Física Teórica y Computacional, Universidad de Granada, 18071 Granada, Spain\looseness=-1}%

\author{Rafael S\'anchez}
\affiliation{Departamento de F\'isica Te\'orica de la Materia Condensada, Universidad Aut\'onoma de Madrid, 28049 Madrid, Spain\looseness=-1}
\affiliation{Condensed Matter Physics Center (IFIMAC), Universidad Aut\'onoma de Madrid, 28049 Madrid, Spain\looseness=-1}
\affiliation{Instituto Nicol\'as Cabrera, Universidad Aut\'onoma de Madrid, 28049 Madrid, Spain\looseness=-1}

\author{Laiachi El Kaoutit}
\affiliation{Departamento de \'Algebra, Universidad de Granada, 18071 Granada, Spain}%

\author{Daniel Manzano}\email[Corresponding author: ]{manzano@onsager.ugr.es}%
\affiliation{Electromagnetism and Condensed Matter Department, Universidad de Granada, 18071 Granada, Spain\looseness=-1}%
\affiliation{Instituto Carlos I de Física Teórica y Computacional, Universidad de Granada, 18071 Granada, Spain\looseness=-1}%

\author{Antonio Lasanta}\email[Corresponding author: ]{alasanta@ugr.es}%
\affiliation{Instituto Carlos I de Física Teórica y Computacional, Universidad de Granada, 18071 Granada, Spain\looseness=-1}%

\affiliation{Departamento de \'Algebra, Facultad de Educaci\'on, Econom\'ia y Tecnolog\'ia de Ceuta, Universidad de Granada, Cortadura del Valle, s/n, 51001 Ceuta, Spain}%

\affiliation{Nanoparticles Trapping Laboratory, Universidad de Granada, Granada, Spain}%

\date{\today}

\begin{abstract}
An intriguing phenomenon in non-equilibrium quantum thermodynamics is the asymmetry of thermal processes. Relaxation to thermal equilibrium is the most important dissipative process, being a key concept for the design of heat engines and refrigerators, contributing to the study of foundational questions of thermodynamics, and being relevant for quantum computing through the process of algorithmic cooling. Despite their importance, the dynamics of these processes are far from being understood. We show that the free relaxation to thermal equilibrium follows intrinsically different paths depending on whether it involves the temperature of the system to increase or to decrease. Our theory is exemplified using the recently developed thermal kinematics based on information geometry theory, utilizing three prototypical examples: a two-level system, the quantum harmonic oscillator, and a trapped quantum Brownian particle, in all cases showing faster heating than cooling under the appropriate conditions. 
A general understanding is obtained based on the spectral decomposition of the Liouvillian and the spectral gap of reciprocal processes.
\end{abstract}

\maketitle

%
%
\section{Introduction}
When a system is pushed far from equilibrium, its evolution may follow anomalous paths. A series of seminal works done during the past century~\cite{Nyquist1928,Zwanzig1965,onsager1931,onsager19312,kubo1957linearresponse,kubo1966,marconi2008} has provided essential advances in studying transitory phenomena in the linear regime associated with fluctuations, except for some particular cases~\cite{evans20081,evans20082} where predictions can extend beyond equilibrium. Despite this progress, we still lack a general theory beyond linear response and fluctuation theorems to decipher the dynamics and behavior of transient regimes of a \textit{freely} evolving system between two desired states~\cite{Seifert2012,strasberg2022quantum}. This problem is of particular interest for quantum information~\cite{weber_mapping_2014,cottet_observing_2017}, quantum thermodynamic processes~\cite{binder:2018,palmero_pre_19,xu_pra_23}, finite-time quantum heat engines~\cite{myers_quantum_2022,liliana_review,cangemi_quantum_2023,tejero2024pre,tejero_preprint} and establishing speed limit bounds \cite{mandelstam1945uncertainty,garcia2022unifying,hamazaki2022speed,nicholson2020time,srivastav2024family}, as well as for transport in interacting nanojunctions~\cite{seoane_transient_2018,heckschen_pair_2022}, where propagation along chiral edge states~\cite{granger_onservation_2009,nam_thermoelectric_2013} can be used to resolve the thermalization process spatially~\cite{altimiras_nonequilibrium_2010,lesueur_energy_2010,ota_spectroscopic_2019,rodriguez_relaxation_2020,braggio_nonlocal_2024}. Recent progress in this direction has been done by unraveling anomalous shortcuts during relaxation processes in out-of-equilibrium systems~\cite{pemartin2024shortcuts}.

A remarkable example of a possible counter-intuitive behavior of a system is the Mpemba-like effect (ME) \cite{jeng2006,mpemba1969,Bechhoefer2021,teza2025speedups}. Namely, put two identical systems at different initial temperatures in contact with a reservoir at a hotter or colder temperature than those of the two systems. The ME occurs when the initially hotter/colder system cools down/heats up faster than the system that was initially closer to the final temperature. In the case of cooling, the effect is called normal ME, and for heating, it is called inverse ME \cite{LuRaz2017,lasanta2017}. In Markovian systems, the ME can be well understood using a spectral decomposition of the decay modes, diminishing (weak ME), or canceling slow-decaying modes (strong ME) to enhance the fast ones, making it possible to control the speed of the relaxation. In this way, up to an exponential acceleration is achievable~\cite{klich2019}. This phenomenon has been realized both in classical~\cite{LuRaz2017,klich2019,kumar2020,walker2021,biswas2023,Schwar2022,kumar22,Vadakkayil2021,pagare2024information,Teza2022, Teza2023_mpemba_boundary, teza2023eigenvalue,Teza2020} and open quantum systems~\cite{carollo2021,nava2019,Chatterjee20232,kochsiek2022accelerating,moroder2024thermodynamics,shapira2024mpemba,strachan2024non,nava2024mpemba,ares2025quantum}. Additionally, a generalization of the ME to quantum entangled configurations has been very recently proposed~\cite{yamashika2024entanglement,joshi2024observing,murciano2024entanglement,rylands2023microscopic}. Note that a strong relation exists between exceptional points  and speed up relaxation in open quantum systems \cite{Chatterjee20232,zhou2023accelerating}.

Alternatively, when spectral methods are not applicable, other strategies can be used to understand anomalous evolution using macroscopic observables depending on the system of interest. The origin of anomalous relaxation is associated with energy non-equipartition in water and granular gases composed by rough hard spheres~\cite{torrente2019,gijon2019}, a particular condition in kurtosis also in the former with smooth hard spheres~\cite{lasanta2017,mompo2021}, and correlation length in spin glasses~\cite{BaityJesi2019}. Furthermore, the strategy of employing several sudden changes in temperature has been proven useful for shortening relaxation times, such as preheating protocols~\cite{gal2020}. This approach takes advantage of the slow growth of magnetic domains near phase transitions in systems where time-scale separation is not possible~\cite{pemartin2021}, or through different control techniques~\cite{dann_shortcut_2019,guery2022,chittari2023}.

\begin{figure}[t]
\includegraphics[width=1\linewidth]{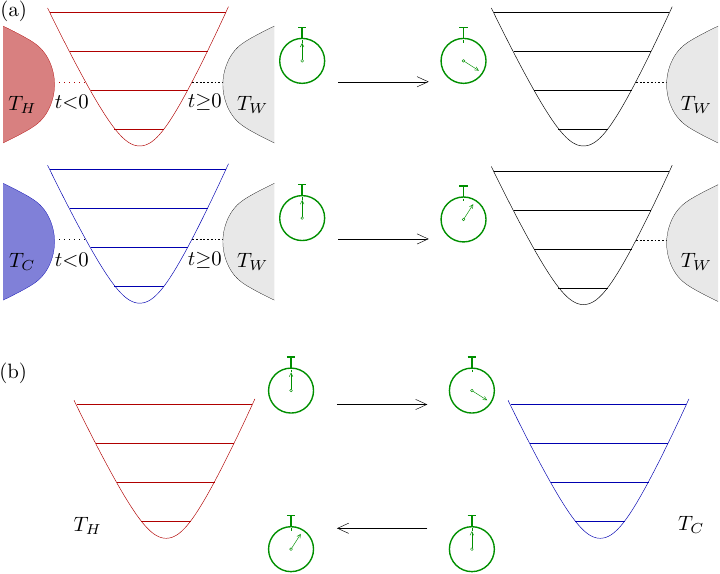}
\caption{\label{fig:scheme}(a) Asymmetric cooling and heating relaxation to an equidistant stationary state at temperature $T_W$. Thermalization takes longer when the system is initially hot (i.e., thermalized with a bath at $T_H$, which is decoupled at $t=0$) than when it is initially cold (at $T_C$), with $T_C<T_W<T_H$. (b) Asymmetric cooling and heating evolution between two states at temperatures $T_C$ and $T_H$, with $T_C<T_H$. The evolution from hot to cold (cooling) is slower than from cold to hot (heating).}
\end{figure}

A fundamental question, illustrated in Fig.~\ref{fig:scheme}, is whether free cooling and heating  processes after a sudden change of the environment temperature are identical or follow intrinsically different paths, see Fig.~\ref{fig:scheme}(a). In classical systems heating and cooling can show an asymmetry that has been verified both theoretically and experimentally far from equilibrium ~\cite{lapolla2020,ibanez2023} \textcolor{black}{and in simple few-level systems~\cite{VanVu2021,manikandan2021}}. 
\textcolor{black}{In Ref.~\cite{VanVu2021}, Vu and Hasegawa use several particular assumptions to demonstrate the non-universality of the asymmetry between thermodynamically equidistant quenches in discrete systems with more than two energy levels. This happens for particular choices of the energy gap and  transition rates.  In this work, we present new and complementary results and new protocols that explore the existence of asymmetry in quantum systems of different nature and complexity.}   An even more emphatic result is that the asymmetry is revealed when relaxation processes occur between two fixed temperatures~\cite{ibanez2023}, see Fig.~\ref{fig:scheme}(b). This has been successfully explained mathematically by using the so-called \emph{thermal kinematics}~\cite{ibanez2023}, based on information geometrical arguments~\cite{crooks2007, Ito2020Stochastic,bravetti2024asymmetric}.  In this paper we focus on that question, that is, unraveling the mechanism of the heating and cooling processes in the realm of open quantum systems. In order to do this we use geometric concepts of quantum information theory~\cite{deffner2017qsl,liu2020quantum} to extend the thermal kinematics theory to the thermodynamics of open quantum systems. We analyze whether a relaxation process far from equilibrium, say from an initially hot to a colder thermal state, is equally fast as its reverse, from the colder to the hotter, and relate it to the properties of the spectral gap \cite{minganti2018spectral,cavina2017slow,mori2020resolving,vznidarivc2015relaxation,haga2021liouvillian,wang2023accelerating}. 
To showcase this, we use simple models based on a thermal qubit, a quantum harmonic oscillator, and a quantum Brownian particle. 

The heat properties of such simple quantum systems have recently become accessible experimentally. Solid-state realizations of qubits coupled to fermionic or bosonic reservoirs allow to control the spectral properties, couplings, and temperatures externally~\cite{giazotto_opportunities_2006,courtois_electronic_2014,pekola_colloquium_2021}. This is the case of quantum dot systems~\cite{vanderwiel_electron_2002,burkard_semiconductor_2023}, which can selectively be (un)connected to different reservoirs with gate voltages~\cite{chida:2017} and whose distribution can be measured via charge detectors~\cite{koski_distribution_2013,hofmann_equilibrium_2016,hofmann_heat_2017,manzano_thermodynamics_2021}, or of superconducting circuits coupled to resistors acting as thermal baths via tunable resonators~\cite{ronzani_tunable_2018,senior_heat_2020,gubaydullin_2022,lu_steady_2022,aamir_engineering_2022,aamir_thermally_2023}. Furthermore,  the qubit state can be monitored~\cite{weber_mapping_2014,campagneIbarcq_observing_2016,cottet_observing_2017,ficheux_dynamics_2018,spiecker_twolevel_2023}.
Improvements in high frequency thermometry even allows us to detect single temperature fluctuations~\cite{gasparinetti:2015}. These ingredients make the detection of relaxation paths in quantum information systems possible.

The recent measurement of asymmetric relaxation of a classical particle in a harmonic trap~\cite{ibanez2023} motivates us to treat this problem from a quantum perspective. To do so, we investigate the thermalization of a quantum Brownian particle, a model that has successfully been applied to describe a plethora of quantum effects, such as quantum dissipation~\cite{Weiss2011,hanggi_fundamental_2005}, harmonic oscillators~\cite{agarwal_brownian_1971}, macroscopic quantum tunneling~\cite{caldeira_influnce_1981,caldeira1983a,caldeira1983b}, metastable states~\cite{hanggi_reaction_1990}, single-electron transistors~\cite{schon_quantum_1990}, the spin-boson problem~\cite{legget_dynamics_1995}, or impurity dynamics in Luttinger liquids~\cite{bonart_from_2012} and ultracold atomic gases~\cite{lampo_bose_2017}. Augmenting the number of degrees of freedom comes with longer relaxation time scales, which favors its detection.  We hence emphasize that understanding the relaxation processes is of importance for quantum thermodynamics and for the physics of driven nanoscale devices~\cite{binder:2018,strasberg2022quantum}, the building up of correlations~\cite{seoane_quench_2020}, and the thermalization of macroscopic quantum states~\cite{brantut_thermoelectric_2013,gallegoMarcos_nonequilibrium_2014}. 

The remainder of this paper is organized as follows. In Secs.~\ref{sec:framework} and \ref{sec:measures}, we present the theoretical framework based on the master equation for open quantum systems and the measures of thermodynamic distances. In Sec.~\ref{sec:protocols}, the definitions of the different protocols are provided. 
We then apply these methods to three different systems. 
In Sec.~\ref{sec:tls} we consider the simplest case of a two level system coupled to a thermal bath, which can be solved analytically.  Then Sec.~\ref{sec:models} considers more complex systems, namely the harmonic oscillator and the quantum Brownian particle. Additionally, this section includes numerical simulations for the non-analytically solvable systems. Section~\ref{sec:spectral} provides a theoretical justification for all the phenomena based on the spectrum of the Liouvillian and the influence of the considered initial state on the evolution. Finally, the conclusions are drawn in Sec.~\ref{sec:conclusions}.

%
%
\section{Theoretical framework: Markovian Open Quantum Systems}\label{sec:framework}

The state of a quantum system, weakly coupled to the environment, is described by its reduced
density matrix $\rho(t)$, whose evolution is governed by the Gorini–Kossakowski–Sudarshan–Lindblad (GKSL) quantum master equation \cite{breuer2002,gardiner2004,lindblad1976,gorini1976,benatti2005,alicki2007,baumgartner2008,manzano2020,landi_rmp_22}
\begin{equation}\label{eqn:master}
\dot{\rho}(t)=\mathcal{L}[\rho(t)],
\end{equation} 
where $\mathcal{L}$ is the Liouvillian superoperator
\begin{equation}
\mathcal{L}[\rho(t)]=-\dfrac{i}{\hbar}[H,\rho(t)]+\sum_{i=1}^{N}\left(L_i \rho(t) L^\dagger_i-\frac{1}{2}\left\{L^\dagger_i L_i,\rho(t)\right\}\right),
\label{eqn:lindblad}
\end{equation}
and $H$ is the Hamiltonian of the system describing its coherent dynamics. The $N$ jump operators $L_i$ describe the dissipative effects due to the presence of an environment. The Liouvillian superoperator $\mathcal{L}$ preserves the trace, i.e. $\Tr\left(\mathcal{L}[\rho(t)]\right)=0$, Hermiticity, i.e., $\left(\mathcal{L}[\rho(t)]\right)^\dagger=\mathcal{L}[\rho^\dagger(t)]$, $\forall \rho(t)$, and complete positivity. 

The general solution to Eq.~(\ref{eqn:master}) can be directly obtained as $\rho(t)=e^{t\mathcal{L}}[\rho(0)]$, where the superoperator $e^{t\mathcal{L}}$ is defined by its power expansion. Assuming the generator to be diagonalizable, one finds the right eigenmatrices, $\Lambda^r_k$, such that 
\begin{equation}
\mathcal{L}[\Lambda^r_k]=\lambda_k \Lambda^r_k .
\label{eqn:right_diag}
\end{equation}
The complex numbers $\lambda_k$ are the eigenvalues of the Liouvillian. Note that, due to the Hermiticity preservation of $\mathcal{L}$, if $\lambda_k$ is a complex eigenvalue, then $\lambda_k^*$ must also be an eigenvalue. For the same reason, one can also show that if $\lambda_k$ is real, then $\Lambda^r_k$ can be chosen to be Hermitian. Associated with the map defined in Eq.~\eqref{eqn:lindblad}, there is a dual map, also called the adjoint Lindblad map, which implements the evolution of observables:
\begin{equation}
\mathcal{L}^{\dagger}[O]=\dfrac{i}{\hbar}[H,O]+\sum_{i=1}^{N} \left(L^\dagger_i O L_i-\frac{1}{2}\left\{O,L^\dagger_i L_i\right\}\right).
\end{equation}
This dual map, $\mathcal{L}^{\dagger}$, is diagonalized by the left eigenmatrices $\Lambda^\ell_k$, 
\begin{equation} 
\mathcal{L}^{\dagger}[\Lambda^\ell_k]=\lambda_k \Lambda^\ell_k .
\label{left-diag}
\end{equation}
The matrices $\Lambda^\ell_k$ are in principle different from the matrices $\Lambda^r_k$ in Eq.~\eqref{eqn:right_diag}. However, $\Lambda^\ell_k$ and $\Lambda^r_k$ still form a bi-orthogonal basis for the space of matrices and can always be defined fulfilling the property $\Tr\left(\Lambda^\ell_k \Lambda^r_h\right)=\delta_{kh}$. 

Since the dynamics generated by $\mathcal{L}$ is completely positive, the eigenvalues of the Liouvillian superoperator all have a non-positive real part, ${\rm Re}\left(\lambda_k\right)\le0$. Furthermore, for bounded systems, Evan's theorem \cite{evans1977} enforces that at least one eigenvalue is zero, $\lambda_1=0$, and this is also the case for many unbounded systems. Assuming that the null eigenvalue is non-degenerate, the asymptotic stationary state of the open quantum system is directly related to its associated eigenmatrix~\cite{manzano:av18,thingna:chaos21},
\begin{equation}
\rho_{\rm ss}=\lim_{t\to\infty}\rho(t) = \Lambda_1^r.
\label{rho_ss}
\end{equation}
Integrating Eq.~(\ref{eqn:lindblad}), the spectral decomposition of $\mathcal{L}$ allows us to write the dynamics of any initial density matrix as 
\begin{equation}\label{eqn:evolution_spectrum} 
\rho(t) = e^{t\mathcal{L}}\left[\rho_0\right]=\Lambda^r_1+\sum_{k=2}^{d^2}e^{t\lambda_k}{\rm Tr}\left(\Lambda^\ell_k \rho_0\right)\Lambda^r_k ,
\end{equation}
where $d$ is the dimension of the Hilbert space of the system. This decomposition shows that the matrices $\Lambda_k^r$ are nothing but the excitation modes of the system, each one characterized by a decay rate $|{\rm Re}(\lambda_k)|$. For long times, the relevant terms are those related to the $\lambda_k$ with the smallest real part in modulus and finite overlap with the initial state. To study the time-evolution of our systems we order the eigenvalues $\lambda_k$ in such a way that $|{\rm Re}\left(\lambda_2\right)|\le |{\rm Re}\left(\lambda_3\right)|\le \ldots \le |{\rm Re}\left(\lambda_m\right)|$. The overlap between the $i-$th eigenmatrix and the initial state, $\rho_0$, is determined by
\begin{equation}\label{eqn:overlap}
    \xi_i = \Tr\left(\Lambda_i^\ell \rho_0\right).
\end{equation}
Note that this term $\xi_i$ is the same as the one appearing in the sum presented in Eq. (\ref{eqn:evolution_spectrum}). This term will provide us with the influence of the Lindbladian, which fixes the temporal evolution, onto the initial state.

\section{Quantum Thermal kinematics: Measures of distance and speed}
\label{sec:measures}
The concept of \emph{thermal kinematics}, established for classical systems recently in Ref.~\cite{ibanez2023}, combines arguments from stochastic thermodynamics with information geometry to analyze the thermodynamical processes \cite{Ito2020Stochastic}. For classical systems, it is possible to define a \emph{statistical distance} \cite{ibanez2023, jordan1997free}, related to the classical Fisher information, $I_{\text{cl}}(t)$, which quantifies the temporal variation of local flows. Therefore, for two time-varying infinitesimal processes, the line element can be defined from the Kullback-Leibler divergence (KLD) of two probability distributions, defined as
\begin{equation}\label{eqn:classical_KL}D_{\text{cl}}\left[P_{\text{cl}}(x,t+dt), P_{\text{cl}}(x,t)\right] = I_{\text{cl}}(t) dt^2 + \mathcal{O}(dt^4),
\end{equation}
which allows us to define a proper statistical distance between two states [see Appendix \ref{sec:appendix}, Eq. (\ref{eqn:classical_fisher_appendix})]. Note that we  denote all classical quantities and variables with the subscript $\mathrm{cl}$. The line element is then defined from Eq. (\ref{eqn:classical_KL}) as
\begin{equation}\label{eqn:classical_dl}
    dl_{\mathrm{cl}} := \sqrt{I_{\text{cl}}(t)} dt.
\end{equation}
where $\sqrt{I_{\text{cl}}(t)}$ can be identified as the \emph{statistical velocity} at a given time $t$, namely
\begin{equation}\label{eqn:classical_v}
    v_{\text{cl}}(t) := \sqrt{I_{\text{cl}}(t)}.
\end{equation}
To study thermal kinematics in the quantum regime, we may use two different measures. The first one will be the \emph{fidelity} between two states (analog to the KLD in the classical case), defined as 
\begin{equation}\label{fidelity}
    F(\rho_1,\rho_2):= \Tr\sqrt{\sqrt{\rho_1}\rho_2 \sqrt{\rho_1}}.
\end{equation}
It measures how \emph{close} two quantum states are in terms of their density matrix. It is  symmetric and invariant under unitary operations. Despite it does not define a metric distance \cite{nielsen_00}, the fidelity allows us to define the so-called Bures distance
 \begin{equation}\label{eqn:app_bures}
    \left[D_B(\rho,\sigma)\right]^2 := 2[1 -  F(\rho,\sigma)],
\end{equation}
which is a statistical distance.
Similarly to the classical case, in our context of thermal relaxation, an infinitesimal statistical line element may be defined as follows~\cite{wooters1981,deffner2017qsl,farre2020bounds,liu2020quantum}
\begin{equation}\label{eqn:bures_quantum}
\left[D_B \left(\rho(t), \rho(t+dt) \right)\right]^2 = \frac{1}{4} \mathcal{I}_Q [\rho(t)]  dt^2 + \mathcal{O}(dt^4)
\end{equation}
with respect to the parameter time, being $\mathcal{I}_Q$ the quantum Fisher information (QFI), defined as 
\begin{equation}\label{eqn:quantum_fischer}
    \mathcal{I}_Q [\rho(t)]:= \text{Tr} \left[L_t^2 \rho(t) \right],
\end{equation}
where $L_t$ is the logarithmic time-derivative operator defined by $\dot{ \rho}(t):= \left[L_t \rho(t) + \rho(t) L_t \right]/2$, see Appendix \ref{sec:appendix}. Now, we have a symmetric and proper  metric distance.  From Eq.~(\ref{eqn:bures_quantum}), we can directly define the line element as
\begin{equation}\label{eqn:line_element}
  dl := \frac{1}{2}\sqrt{\mathcal{I}_Q [\rho(t)] }dt, 
\end{equation}
and thus
\begin{equation}\label{eqn:speed}
    v(t) := \frac{1}{2}\sqrt{\textcolor{black}{\mathcal{I}_Q} [\rho(t)]}
\end{equation}
represents the \emph{quantum instantaneous
statistical velocity} of the system in the quantum case \cite{deffner2017qsl,farre2020bounds}. The statistical length of a path taken between time $t_i$ and $t_f$ is computed as 
\begin{equation}\label{eqn:length}
    \ell(t_i,t_f)=\frac{1}{2}\int_{t_i}^{t_f}   \sqrt{\textcolor{black}{\mathcal{I}_Q} \left[ \rho(t) \right]} dt.
\end{equation}
As reaching the steady state during a dissipative process takes infinite time, to establish
a kinematic basis for quantifying thermal relaxation kinematics, we define the  \emph{ quantum degree of completion} as
\begin{equation}\label{eqn:degree}
    \varphi(s) := \dfrac{\ell(t_i,t_s)}{\ell(t_i,t_f)},
\end{equation}
being a monotonically increasing function bounded between 0 and 1.

\section{Heating and cooling protocols} \label{sec:protocols}

To puzzle out the properties of cooling and heating far from equilibrium in quantum systems subject to instantaneous quenches, we define two possible experiments.

\subsection{Three-temperature protocol}\label{subsec:three_temps}
The first feasible protocol is to compare the free evolution with respect to an intermediate temperature. Hence, we define three temperatures $T_C < T_W < T_H$, the subscripts corresponding to \textit{cold} ($C$), \textit{warm} ($W$), and \textit{hot} ($H$) respectively. Associated to this temperatures there are three Gibbs states, $\rho^{\text{th}}_{\beta_i} = \exp[-\beta_i H]/\mathcal{Z}$, 
with $H$ being the Hamiltonian of the system, $\beta_i=1/k_B T_i$ the inverse temperature, and $\mathcal{Z} = \Tr \left\{\exp[-\beta_i H]\right\}$ the partition function for $i \in \{C,W,H\}$.

\textcolor{black}{In this protocol, the system is initially prepared to be thermalized by interacting with either a hot ($\rho^{\text{th}}_{\beta_H}$) or cold bath ($\rho^{\text{th}}_{\beta_C}$). At $t=0$, we introduce a sudden quench by coupling the system to the warm bath, as illustrated in Fig.~\ref{fig:scheme}(a).  
As both trajectories, cooling and heating up, evolve towards the same steady state ($\rho^{\text{th}}_{\beta_W}$), we can use the fidelity between our time-dependent state and the target one as a measure of distance. To fix the initial conditions,  we consider thermal states with equal fidelity values with respect to $T_W$ for both $T_C$ and $T_H$, meaning that 
\begin{equation}
F(\rho^{\text{th}}_{\beta_C},\rho^{\text{th}}_{\beta_W}) = F(\rho^{\text{th}}_{\beta_H},\rho^{\text{th}}_{\beta_W}).
\end{equation}
The relaxation of the heating and cooling processes is then monitored by the evolution of their fidelities. 
}


We first focus on what we call \textit{forward protocol} where the relaxation occurs toward the \emph{warm} temperature, $T_W$ starting from the states at hot, $T_H$, and cold,  $T_C$, temperatures. 
\textcolor{black}{To sort out interpretations related to the different temperature differences for equidistant states (in fact, $T_H-T_W\neq T_W-T_C$), we introduce a {\it backward protocol}:}
we prepare the system to be in equilibrium at the \emph{warm} temperature $T_W$ and track back the relaxations at $T_C$ and $T_H$, respectively. 

\subsection{Two-temperature protocol}
We can also proceed using a simpler protocol, namely, cooling and heating between two temperatures $T_C < T_H$, respectively, as sketched in Fig.~\ref{fig:scheme}(b). 
\textcolor{black}{This protocol eliminates possible effects related to the details different relaxations paths, e.g., that one of them takes place at lower temperatures than the other one.}
In this case the absence of a reference density matrix prevents us to use the fidelity as a distance measure.  We need to use a \textit{true} metric distance, namely, the quantum Fisher information, Eq. (\ref{eqn:quantum_fischer}). 

In this scenario, starting with the system thermalized at one of the temperatures, after a sudden quench, we put the system in contact with a bath at the other temperature and let the system evolve freely. This phenomenon allows us to observe heating, i.e relaxation at $T_H$ in a temperature quench from an equilibrium
prepared at $T_C$; and cooling, i.e. relaxation at $T_C$ in a temperature quench from the
equilibrium at $T_H$. In order to compare the two processes in a proper way we will use the \emph{quantum degree of completion} given by Eq. (\ref{eqn:degree}) and the \emph{quantum instantaneous statistical velocity}, Eq. (\ref{eqn:speed}). 

In the following, we test these protocols in three different quantum systems of increasing complexity. The first one is the simplest case: a two-level system coupled to a thermal bath at a given temperature.  In this case, all the relevant quantities will be obtained analytically, since the solution for the Lindblad master equation is  available exactly. This model will serve as a motivating case to perform an in-depth analysis of the two main models presented in the paper: the quantum harmonic oscillator and a quantum Brownian particle. For the harmonic oscillator, since the Hilbert space characterizing the system has infinite dimensions, all the computations performed are potentially more complicated. For this reason, only some of the results are obtained analytically. Finally, the results presented for a quantum Brownian particle are obtained numerically.

\section{A simple case: Thermal qubit} \label{sec:tls}
Let us start with a preliminary analysis of the simplest system of interest: a two-level  system weakly coupled to a thermal bath. Despite its simplicity, this is a paradigmatic example as the coupling of few-level systems to thermal baths has been mastered in the last decades in different condensed matter platforms, e.g., semiconductor quantum dots~\cite{vanderwiel_electron_2002,burkard_semiconductor_2023} or superconducting qubits~\cite{giazotto_opportunities_2006,courtois_electronic_2014,pekola_colloquium_2021}. They are important pieces in the development of modern quantum thermodynamic  engines~\cite{benenti:2017,myers_quantum_2022,liliana_review,cangemi_quantum_2023}. This simple case provides us with analytical understanding of the problem. It is important to remark that all the final conclusions drawn for the more involved examples of the  harmonic oscillator and quantum Brownian motion will be in accordance with the ones obtained from this simple analysis.

Consider a two-level system weakly coupled to a thermal bath at inverse temperature $\beta$. Transitions between the ground ($n=0$) and the excited ($n=1$) states, split by an energy $\hbar\omega$, occur with rates $W_{10}=\gamma \bar{n}(\omega,T)$ and $W_{01}=\gamma[1+\bar{n}(\omega,T)]$ induced by the bath~\cite{breuer2002}, with the coupling rate $\gamma$ and an average number of photons with frequency $\omega$ in a bath at temperature $T$, $\textcolor{black}{\bar{n}}(\omega,T)$, given by the Bose-Einstein distribution 
\begin{eqnarray}
\label{eqn:bose}
\bar{n}(\omega,T)= \left[\exp(\hbar \omega/k_BT) - 1\right]^{-1},
\end{eqnarray}
where $\hbar$ and $\kB$ are Planck's  and Boltzmann's constants. When thermalized, the state of the system can be written as a vector formed by the diagonal elements of the density matrix giving the occupation of the two states, $\rho=(\rho_{00} \ \rho_{11})^T$ in the Fock-Liouville representation, with $\rho_{00}+\rho_{11}=1$. 
In this case, the Lindblad equation, Eq. (\ref{eqn:lindblad}), is a simple rate equation 
\begin{eqnarray}\label{mat}
\displaystyle
{
\dot\rho(t)=
\left(\begin{array}{cc}
-\gamma \bar{n}(\omega,T) & \gamma[\bar{n}(\omega,T)+1]\\
\gamma \bar{n}(\omega,T) & \textcolor{black}{-}\gamma[\bar{n}(\omega,T)+1]
\end{array}  \right) \rho(t).
}
\end{eqnarray}
Note that, in the absence of coherence in the initial state, the Hamiltonian term of the Lindblad equation~\eqref{eqn:lindblad} does not contribute to the occupations, so the dynamics is purely dissipative. 

We are interested in the relaxation from an initial thermal state at temperature $T_0=T+\Delta T=1/\kB\beta_0$. The time evolution of the density matrix can be obtained solving Eq.~\eqref{mat}
\begin{eqnarray}\label{eqn:rhoev}
\rho(t)&=&\rho_\beta^{\rm th}+\frac{e^{-\Gamma t}(e^{\hbar\omega\beta_0}-e^{\hbar\omega\beta})}{(1+e^{\hbar\omega\beta})(1+e^{\hbar\omega\beta_0})}
\left(\begin{array}{c}
1\\
-1
\end{array}  \right) ,
\end{eqnarray}
with the total rate $\Gamma=\gamma[1+2\bar{n}(\omega,T)]=\gamma\coth(\hbar\omega\beta/2)$, that is proportional to the thermal fluctuations of the bath. 
The fact that there is a single relaxation channel in this case 
\textcolor{black}{makes it clear that the hotter the steady-state bath, the larger the relaxation rate of the system will be.}

\begin{figure}[t]
\includegraphics[width=0.7\linewidth]{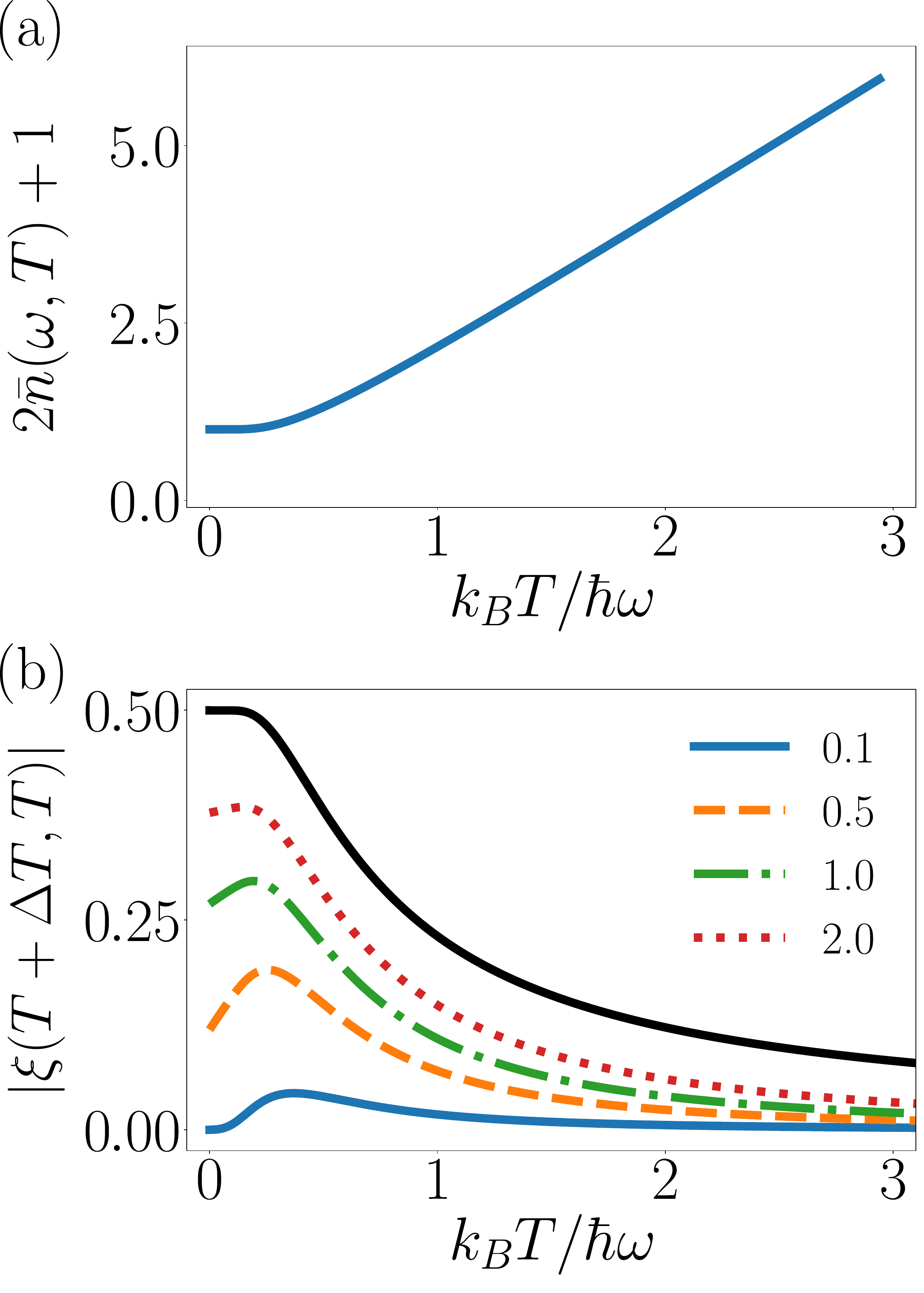}
\caption{\label{fig:plots_qubit} Thermalization kinematics for a qubit. (a) Dependence of the decaying mode corresponding to the non-zero eigenvalue. (b) Overlap \textcolor{black}{of the first decaying mode of a} state thermalized at a temperature $T+\Delta T$ with the stationary state at a temperature $T_W$, for different values of $\Delta T/T$. The black line in (b) corresponds to the asymptotic behavior at large $\Delta T$.
} 
\end{figure}

It is, however, convenient to look further into the details of the dynamics, as introduced in Sec. \ref{sec:framework}. We start by obtaining the eigenvalues of $\mathcal{L}$. In this simple case, the spectrum is reduced to only two values: $\lambda_1=0$ related to the trivial stationary state, and $\lambda_2=-\Gamma$, which takes into account the decay mode and depends on the bath parameters, contained in $\bar{n}(\omega,T)$, and on the coupling $\gamma$. The temperature dependence of the decaying mode corresponding to $\lambda_2$ is plotted in Fig.~\ref{fig:plots_qubit}(a). Their corresponding (right) eigenvectors, see Eq.~\eqref{left-diag}, are given by 
\begin{eqnarray}
\label{eq:L1r}
\displaystyle
{
\Lambda_1^r=\rho_\beta^{\rm th}=\frac{1}{e^{\beta\hbar\omega}+1}
\left(\begin{array}{c}
e^{\beta\hbar\omega}\\
1
\end{array}  \right),
}
\end{eqnarray}
corresponding to the stationary state of the system, characterized by $\lambda_1$, and
\begin{eqnarray}
\displaystyle
{
\Lambda_2^r=
\left(\begin{array}{c}
1\\
-1
\end{array}  \right),
}
\end{eqnarray}
which is related to the decaying mode, $\lambda_2$.

These results agree with the exact evolution obtained in Eq.~\eqref{eqn:rhoev}, where we identify
\begin{equation}
\xi\equiv\xi_2=\frac{e^{\beta_0\hbar\omega}-e^{\beta\hbar\omega}}{\left(e^{\beta_0\hbar\omega}+1\right)\left(e^{\beta\hbar\omega}+1\right)}.
\end{equation}
as the overlap $\xi(T+\Delta T,T)\equiv\Tr(\Lambda_2^\ell\rho_0)$ between the initial state $\rho_0=\Lambda_1^r(T+\Delta T)$ (a thermal distribution at temperature $T_0$) and the decaying mode 
\begin{equation}
\Lambda_2^\ell=\frac{1}{1+e^{\hbar\omega\beta}}(1 \ -e^{\hbar\omega\beta}),
\end{equation}
see Eq.~(\ref{eqn:overlap}). \textcolor{black}{Note that, as expected from a nonequilibrium quantity, the overlap vanishes in equilibrium $[\xi(T,T)=0]$, and it's modulus is invariant under the exchange of temperatures $T$ and $T_0$ }
\beq
\label{eq:xisym}
|\xi(T_0,T)|=|\xi(T,T_0)|.
\eeq
As shown in Fig.~\ref{fig:plots_qubit}(b), the overlap increases monotonically with $\Delta T$, i.e., far from equilibrium states are more strongly overlapped \textcolor{black}{with the decaying mode}.
The overlap of an infinite-temperature and a zero-temperature states is maximal: $|\xi|\to1/2$, with the bound $|\xi(T+\Delta T,T)|\leq |\xi_{\rm asym}|$ for the asymptotic value
\begin{equation}
|\xi_{\rm asym}|\to\frac{1}{2}\tanh\left(\frac{\hbar\omega}{2k_BT}\right),
\end{equation}
when $\Delta T\gg T$, see black curve in Fig.~\ref{fig:plots_qubit}(b).
This overlap is a measure of the speed of decaying to the new steady state of the system and gives information on the influence of the initial state in the dynamics.
\textcolor{black}{In this single-mode problem, where the density matrix can be expressed in terms of a single occupation, say that of the ground state, $\rho_{00}(t)=\rho^{\textrm{th}}_{\beta,00}+\xi e^{-\Gamma t}$, c.f. Eq.~\eqref{eqn:rhoev}, the overlap can be interpreted as the total change in the qubit populations between the initial and the stationary states: $\xi=\rho^{\textrm{th}}_{\beta,00}-\rho_{00}(0)=-[\rho^{\textrm{th}}_{\beta,11}-\rho_{11}(0)]$. }
\textcolor{black}{In the relaxation from an initial state with overlap $\xi$, a given difference of population $\delta\rho\equiv\rho_{00}(\tilde t)-\rho^{\textrm{th}}_{\beta,00}$ will be attained in a time 
\begin{equation}
\label{eq:qubit_dt}
\tilde t=\frac{1}{\Gamma}\ln(\xi/\delta\rho). 
\end{equation}
Hence, for two states decaying toward the same stationary state (i.e., with the same rate $\Gamma$), the one with a larger overlap will take a longer time $\tilde t$.}

\begin{figure}[t]
\includegraphics[width=.75\linewidth]{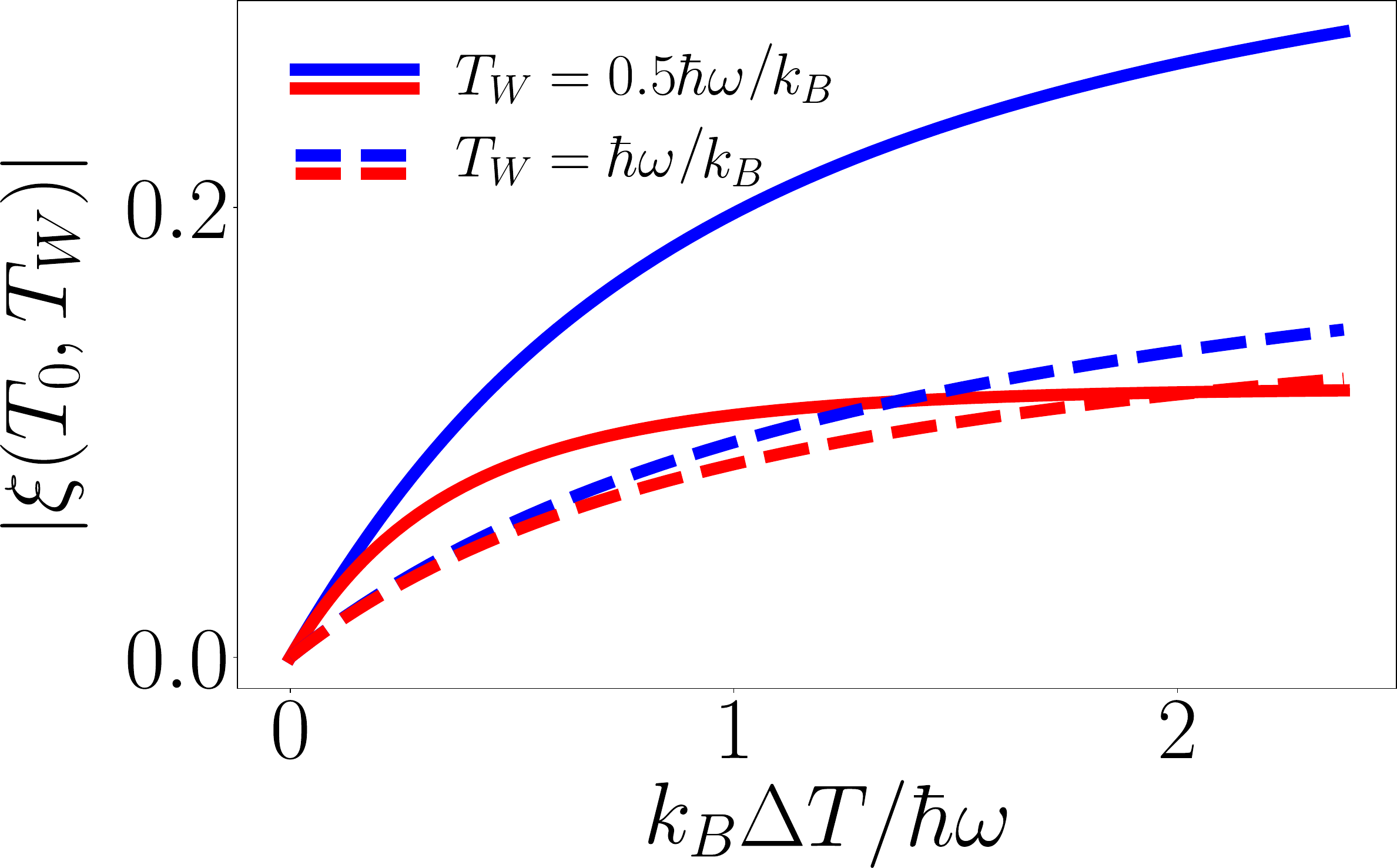}
\caption{\label{fig:qubit_ovrlp}Overlap \textcolor{black}{of the first decaying mode of the cooling ($T_0=T_H$, blue) and heating ($T_0=T_C$, red lines) processes after quenches from two equidistant states with respect to a thermal distribution a temperature $T_W$. $\Delta T$ is the increase in temperature $T_H=T_W+\Delta T$. The cold temperature $T_C$ is chosen accordingly.} 
} 
\end{figure}
These properties already \textcolor{black}{shed} some light on the behavior of the two protocols: the symmetry \eqref{eq:xisym},  involves that in a two-temperature protocol, the overlap is the same in both ways (cooling and heating), $|\xi(T_H,T_C)|=|\xi(T_C,T_H)|$: therefore any asymmetry in such protocol is to be attributed only to the monotonous increase of the coupling rate shown in Fig.~\ref{fig:plots_qubit}(a): indeed \textcolor{black}{$\Gamma$} is larger when relaxing to a hot bath, so $|\lambda_2^{C{\to}H}|>|\lambda_2^{H{\to}C}|$. 
However, for a three-temperature protocol we have the opposite situation: $|\lambda_2^{C{\to}W}|=|\lambda_2^{H{\to}W}|$, so an asymmetric relaxation is to be attributed to the \textcolor{black}{temperature dependence of the overlap with the decaying mode. As shown in Fig.~\ref{fig:qubit_ovrlp}, we always find $|\xi(T_H,T_W)|>|\xi(T_C,T_W)|$ for pairs of Bures-equidistant initial states. 
Therefore, according to Eq.~\eqref{eq:qubit_dt}, cooling is expected to be a slower process.}
This analysis shows that, even in this simple configuration, the two- and three-temperature protocols are intrinsically different: while the former can be understood by the higher fluctuations of a hot bath \textcolor{black}{governing the decaying rate}, \textcolor{black}{the later relies on the different overlap of the initial states with the decaying mode.} 

\begin{figure*}[t]
\includegraphics[width=0.75\linewidth]{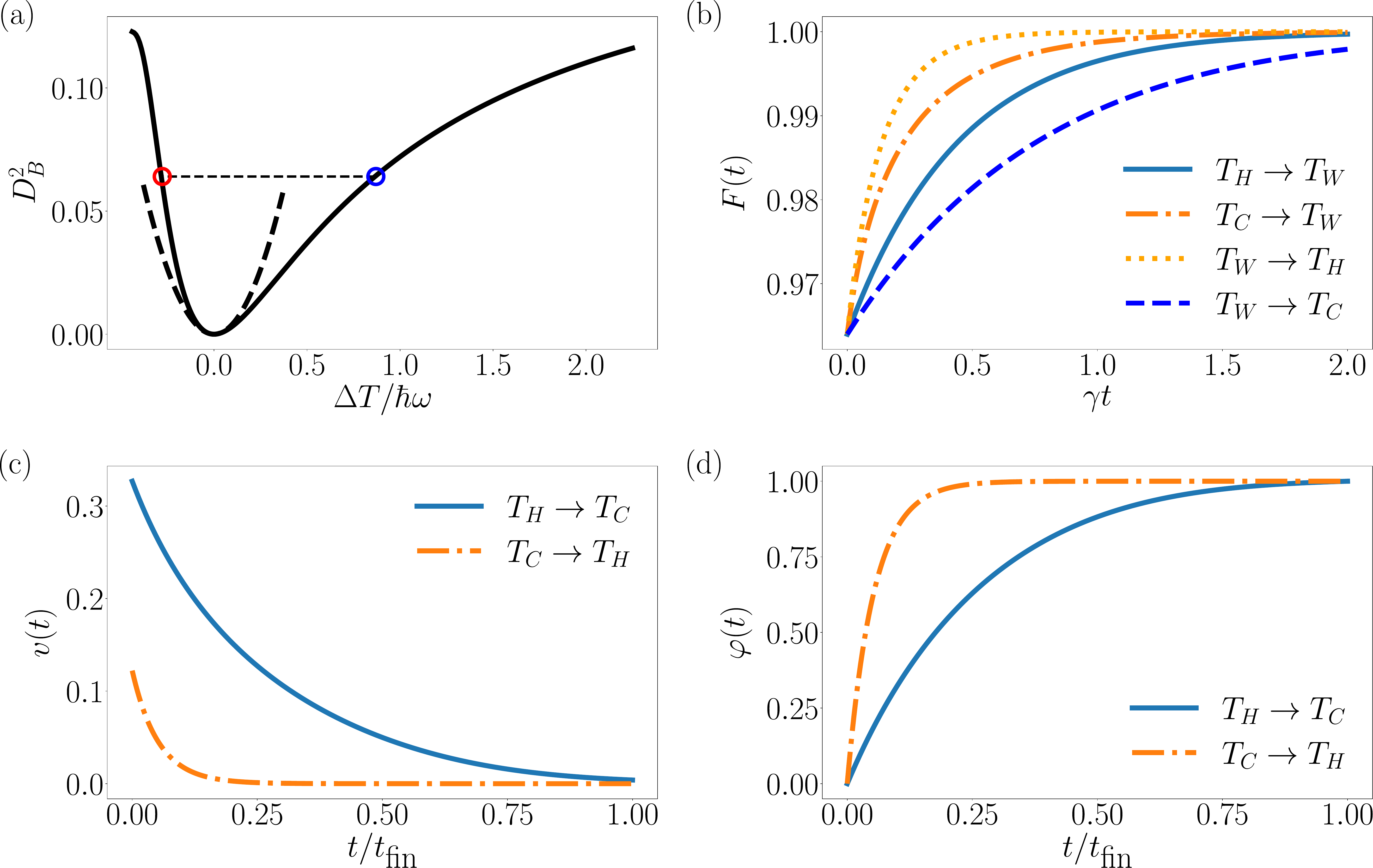}
\caption{\label{fig:panel_qubit}\small Relaxation of a thermal qubit. (a) Bures distance as a function of $\Delta T$. The blue and red dots mark the equidistant temperatures $T_H$ and $T_C$ considered for the cooling and heating protocols, respectively. These are used to compute (b) the fidelity in the direct and inverse three-temperature protocol (from $T_H$ and $T_C$ to $T_W$ and the opposite). (c) Velocity as function of time, and (d) level of completion for a given time normalized by $t_{\rm fin}=6/\gamma$ in the two terminal protocol between $T_H$ and $T_C$. In all panels   $\hbar\omega=1$, $T_W=\hbar\omega/2\kB$, and $\Delta T=T_H-T_W=\hbar\omega/\kB$, with $T_C\approx\textcolor{black}{0.30\hbar\omega/\kB}$ chosen to be equidistant from $T_W$ as depicted in (a) and $t_0=0$. The dashed line in (a) is a quadratic expansion of $D_B^2$ around $\Delta T=0$, according to Eq.~\eqref{eqn:QBF0exp}. 
}
\end{figure*}

To compute the fidelity as a measure of distance between two thermal states, let the system be described by a Gibbs state at an inverse temperature $\beta$ and frequency $\omega$ in the Fock-Liouville space, $\rho^{\text{th}}_\beta$ given by Eq.~\eqref{eq:L1r}.
Consider two thermal states at different inverse temperatures $\beta_1$ and $\beta_2$. In this case, the fidelity is simply
\begin{equation}
\label{eqn:QBf0}
F(\rho_{\beta_1}^{\rm th},\rho_{\beta_2}^{\rm th})=\frac{1+e^{\hbar \omega(\beta_1+\beta_2)/2}}{[(1+e^{\beta_1\hbar\omega})(1+e^{\beta_2\hbar\omega})]^{1/2}}.
\end{equation}
With this expression, we calculate the Bures distance $D^2_B=2[1-F(\rho_{\beta_1}^{\rm th},\rho_{\beta_2}^{\rm th})]$, see Eq.~(\ref{eqn:app_bures}) and Fig.~\ref{fig:panel_qubit}(a). 

For the density matrix evolution given by  Eq.~\eqref{eqn:rhoev}, we find an analytical expression for the time evolution of the fidelity of a system initially at a state $\rho_{\beta_0}^{\rm th}$ with respect to the stationary state as it is put in contact with a bath at inverse temperature $\beta$, namely
\begin{equation}
\label{eqn:QBft}
F[\rho(t),\rho_\beta^{\rm th}]=\frac{\sqrt{e^{\hbar\omega\beta}[e^{\hbar\omega\beta_0}+A_{\beta\beta_0}(t)]}+\sqrt{1-A_{\beta\beta_0}(t)}}{\textcolor{black}{[(1+e^{\hbar\omega\beta})(1+e^{\hbar\omega\beta_0})]^{1/2}}},
\end{equation}
where the time dependence is encapsulated in the term $A_{\beta\beta_0}(t)\equiv{(1-e^{-\Gamma t}})(e^{\beta\hbar\omega}-e^{\beta_0\hbar\omega})/(1+e^{\beta\hbar\omega})$. Note that at time $t=0$, \textcolor{black}{$A_{\beta\beta_0}(0)=0$}, thereby recovering the fidelity given by Eq.~\eqref{eqn:QBf0}. The fidelity as a function of time is plotted in Fig.~\ref{fig:panel_qubit}(b) for the heating up and cooling down processes in a three-temperature protocol, this result confirming that the heating protocol is faster than the cooling one even when both correspond to a bath at the same temperature $T=T_W$, as anticipated by its smaller overlap shown in Fig.~\ref{fig:qubit_ovrlp}. We have verified that this effect is much stronger for low temperatures. \textcolor{black}{Remarkably, reversing the protocol, i.e., quenching the initial state at $T_W$ to couple it to baths at temperatures $T_H$ and $T_C$, we also find that the heating processes $T_W\to T_H$ and $T_C\to T_W$ are faster than the respective cooling ones ($T_H\to T_W$ and $T_W\to T_C$), see Fig.~\ref{fig:panel_qubit}(b). Note that these four processes realize two two-temperature protocols, confirming that in those the heating process is always faster.}

We get additional insight by using the quantum Fisher information to compute the thermal kinematic distance and velocity for the two-temperature protocols, see Eqs.~(\ref{eqn:line_element}) and (\ref{eqn:speed}). 
In this case, being the density matrix diagonal, and $\partial_t\rho_{00}(t)=-\partial_t\rho_{11}(t)$, the QFI reads ${\cal I}_Q[\rho(t)]=[\partial_t\rho_{00}(t)]^2/\rho_{00}(t)\rho_{11}(t)$, where we have also used $\rho_{00}(t)+\rho_{11}(t)=1$, leading to
\begin{equation}
\label{eq:tls_fisher}
{\mathcal{I}}_Q[\rho(t)]=\frac{\Gamma^2}{[e^{\hbar\omega\beta}\kappa(t)-1][\kappa(t)+1]},
\end{equation}
where the time-dependence is encapsulated in the term 
\begin{equation}
\kappa(t)\equiv-e^{\Gamma t}/(e^{\hbar\omega\beta}+1)\xi.
\end{equation}
\textcolor{black}{Equation \eqref{eq:tls_fisher} shows explicitly that ${\cal I}_Q$ is asymmetric under the exchange $\beta\leftrightarrow\beta_0$.}
With this, we write the instantaneous statistical velocity
\begin{equation}
v(t)=\frac{\Gamma/2}{\sqrt{e^{\hbar\omega\beta}\kappa(t)-1}\sqrt{\kappa(t)+1}},
\end{equation}
with the property that the velocity is low for states with similar temperatures: $v\to ce^{-\Gamma t}$ with $c\ll1$, when $\beta_0\to\beta$. As shown in Fig.~\ref{fig:panel_qubit}(c), though the velocity of the cooling process is larger, the faster decay of $v(t)$ for the heating mechanism indicates that it approaches the thermal state much earlier. To analyze the full process, we compute the statistical length
\begin{gather}
\begin{aligned}
\ell(t_0,t)&=\frac{1}{\Gamma}\left\{\arctan\left[\left|\frac{\left(e^{\hbar\omega\beta}{-}1\right)\kappa(t)-2}{2\sqrt{[e^{\hbar\omega\beta}\kappa(t){-}1][\kappa(t){+}1]}}\right|\right]\right.\\
&-\left.\arctan\left[\left|\frac{\left(e^{\hbar\omega\beta}-1\right)\kappa(t_0)-2}{2\sqrt{[e^{\hbar\omega\beta}\kappa(t_0){-}1][\kappa(t_0){+}1]}}\right|\right]\right\},
\end{aligned}
\end{gather}
which we use to plot the ratio $\varphi(t)=\ell(t_0,t)/\ell(t_0,t_{\rm fin})$ in Fig.~\ref{fig:panel_qubit}(d). It confirms that the heating protocol is indeed faster, despite having a smaller velocity, cf. Fig.~\ref{fig:panel_qubit}(c), as expected for it having a larger $|\lambda_2|$.

\begin{figure*}[t]
\centering
\includegraphics[width=0.75\linewidth]{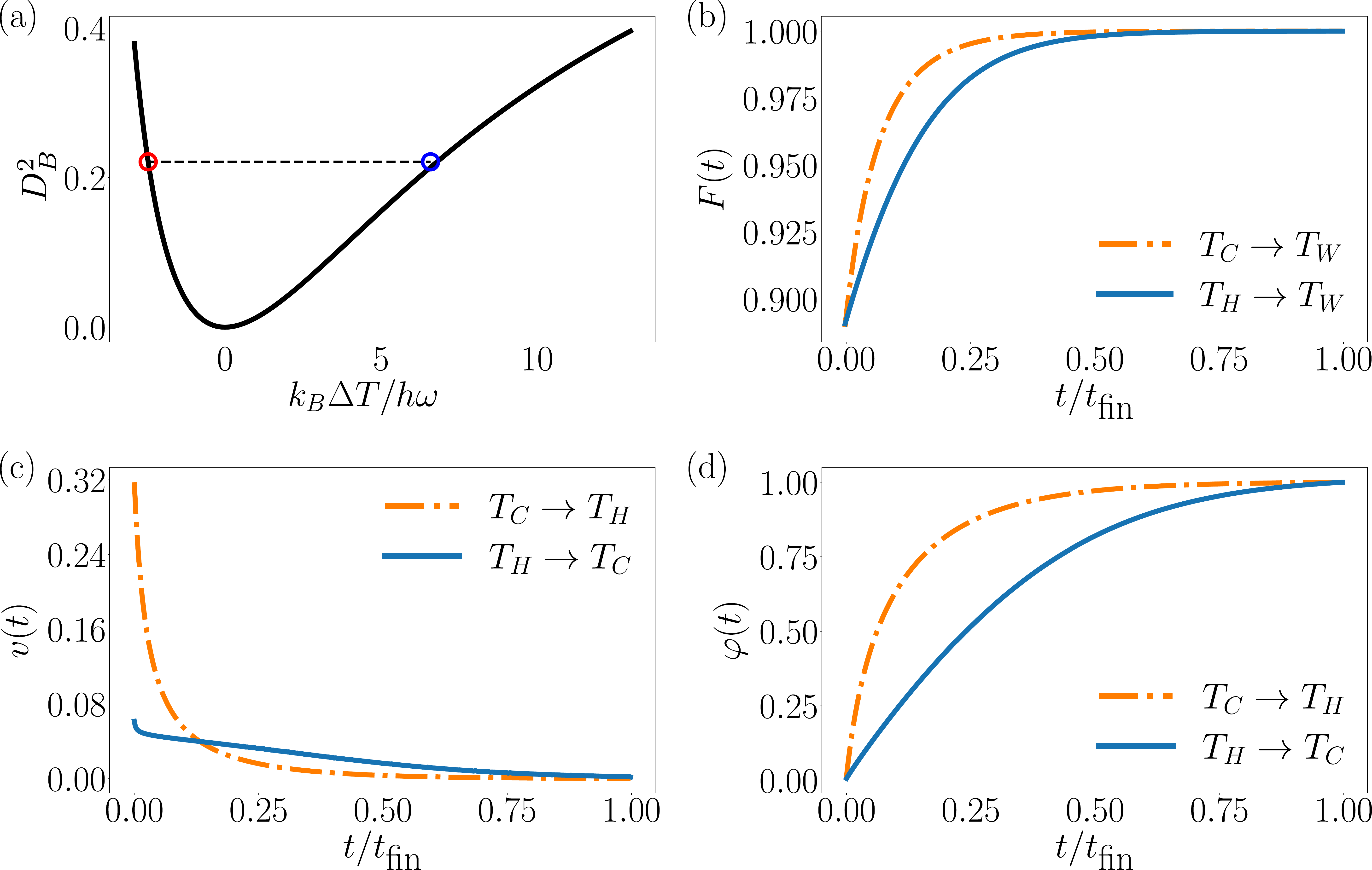}
\caption{Simulation of the protocols for the harmonic oscillator. (a) Bures distance as a function to the temperature difference with respect to the equilibrium state at $T_W$. The red and blue dots mark the initial temperatures $T_C$ and $T_H$ for the heating and cooling processes, respectively. (b) Fidelity with respect to the thermal state at $T_W$ corresponding to the heating (orange) and cooling (blue line) processes of the three-temperature protocol. (c) Instantaneous statistical velocity and (d) degree of completion computed for the heating and cooling processes in the two-temperature protocol. 
In all plots, the temperature ranges are such that $n(\omega,T_C) = 1$ and $n(\omega,T_H) = 10$, with $\gamma = 0.1$, $\hbar \omega = 1$, and $t_{\textrm{fin}} = 50$.}
\label{fig:plots_HO}
\end{figure*}

\section{Increasing complexity}\label{sec:models}

\subsection{Quantum harmonic oscillator}
After having introduced the main concepts presented in the paper with a clear and analytically solvable case, we will perform a similar analysis for more complicated and richer systems. The quantum harmonic oscillator allows us to derive some analytical expressions for the behavior of the system, however, numerical methods are required to compute quantities of interest such as the quantum speed through the quantum Fisher information.

A harmonic oscillator is described by the Hamiltonian
\begin{eqnarray}\label{eqn:model_HO}
    H &=& \hbar \omega a^{\dagger}a,
\end{eqnarray}
where $\omega$ is the oscillator frequency and $a,a^\dagger$ are the annihilation and creation bosonic operators, respectively. The interaction with a thermal bath is described by the jump operators 
\begin{equation}
\label{eqn:model_HO_jump}
    L_+{=} \sqrt{\gamma \bar{n}(\omega,T)} \;a^\dagger \quad{\rm and}\quad
    L_-{=}\sqrt{\gamma[\bar{n}(\omega,T){+}1]} \;a,
\end{equation}
where $\gamma$ is the coupling strength with the bath, and $\bar{n}(\omega, T)$ is the average number of excitations in the bath at a given temperature $T$, see  Eq.~(\ref{eqn:bose}).  The state of a thermal harmonic oscillator and its dependence on temperature and frequency are determined by $\bar{n}(\omega,T)$ as
\begin{gather}
\begin{aligned}
\label{eqn:thermal_HO}
    \rho^{\text{th}}_\beta &= 
    \sum_{n=0}^\infty \dfrac{\left[\bar{n}((\omega,T)\right]^n}{[1+\bar{n}((\omega,T)]^{n+1}} \ket{n}\bra{n} \\
    &=\textcolor{black}{2\sum_{n=0}^\infty e^{-(n+1/2)\hbar\omega\beta}\sinh\left(\frac{\hbar\omega\beta}{2}\right) \ket{n}\bra{n},}
\end{aligned}
\end{gather}
being $\ket{n}$ the pure state of a system with $n$ photons. 
Initially, we will consider the system to be in such thermal state.

If a system is in a Gaussian state, including a thermal state, and the interaction with the bath is also Gaussian, its state would be entirely characterized by the evolution of its occupation numbers $\left< a \right>$ and $\left\langle a^{\dagger} a\right\rangle$. This fact reduces the problem to the computation of the evolution of the expected values instead of the whole density matrix, leading to a single ordinary differential equation. The dynamics of $\langle a\rangle$ and $\left\langle a^{\dagger} a\right\rangle$ are described by the following expressions~\cite{asadian:pre13}:
\begin{eqnarray}
 \frac{d\langle a\rangle}{d t} & = & -i\left(\omega+\dfrac{\Gamma}{2}\right)\langle a\rangle, \nonumber \\
\frac{d\left\langle a^{\dagger} a\right\rangle}{d t} & = & -\Gamma\left\langle a^{\dagger}a\right\rangle+\Gamma \bar{n}(\omega,T).
\label{eqn:fp}
\end{eqnarray}   
Focusing on the temporal evolution of the average number of the system excitations, the solution to this differential equation can be obtained, leading us to the variation in the average number of excitations
\begin{align}\label{eqn:ad_a_temp}
\left<a^\dagger a  \right>_t = & \left<a^\dagger a \right>_0 e^{-\Gamma t} + \int_0^t \Gamma n(\omega,T)  e^{-\Gamma (t-s) } ds,
\end{align}
Note that the sub-index $t$ represents the time dependence and $0$ the initial value for the average number of excitations. We  consider that the system and the bath are in contact at $t=0$, without loss of generality. The protocol is modeled by a quench, i.e. a step function, so $\bar{n}(\omega, T)$ is constant within the integral. Thus

\begin{equation}
    \left<a^\dagger a  \right>_t =  \left<a^\dagger a \right>_0 e^{-\Gamma t} + n(\omega,T)\left(1-e^{-\Gamma t}  \right).
\end{equation}
%
With this, we compute analytically the time evolution of the fidelity with respect to the stationary state, which reads
\begin{widetext}
\begin{align}\label{eqn:theoretical_F}
    F(\rho^{\text{th}}_S(t), \rho_{\text{F}}) =  \Tr \sum_{n=0}^{\infty}\left[ \dfrac{\left(\left<a^\dagger a  \right>_t \bar{n}(\omega,T)\right)^{n}}{(1+\left<a^\dagger a  \right>_t)^{1+n}(1+\bar{n}(\omega,T))^{1+n}}\right]^{1/2} \ket{n}\bra{n} 
    &= \dfrac{1}{[(1+\left<a^\dagger a  \right>_t)(1+\bar{n}(\omega,T))]^{1/2}} \dfrac{1}{1-r},
\end{align}
where $r = \left\{\left<a^\dagger a  \right>_t \bar{n}(\omega,T)/[(1+\left<a^\dagger a  \right>_t)(1+\bar{n}(\omega,T))]\right\}^{1/2}$.
\end{widetext}
In Fig.~\ref{fig:plots_HO} both the Bures distance as a function of the temperature, Fig. \ref{fig:plots_HO}(a), and the fidelity as a function of time, Fig. \ref{fig:plots_HO}(b), are displayed. The results confirm behavior obtained for the two-level system. One main difference is that to obtain the same Bures distance, we need a higher temperature difference in the harmonic oscillator case than for the two-level system. This is due to the infinite size of the Hilbert space of the harmonic oscillator, in comparison to a two-dimensional Hilbert space. 

To analyze the thermal kinematics of the system we have computed numerically the quantum Fisher information of the two-temperature protocol. The results are displayed in Figs.~\ref{fig:plots_HO}(c) and \ref{fig:plots_HO}(d). 
It is clear that, even if the harmonic oscillator is a different and more complicated system, its thermal behavior is similar to the one for the two-level system. 
\textcolor{black}{Note that the velocities behave differently: despite $v$ being initially larger for the heating process, it has a faster decay, again signaling that the stationary state is reached earlier. }
In the next section we check this behavior with an even more complex system as the quantum Brownian particle.

\begin{figure*}[t]
\centering
\includegraphics[width=0.75\textwidth]{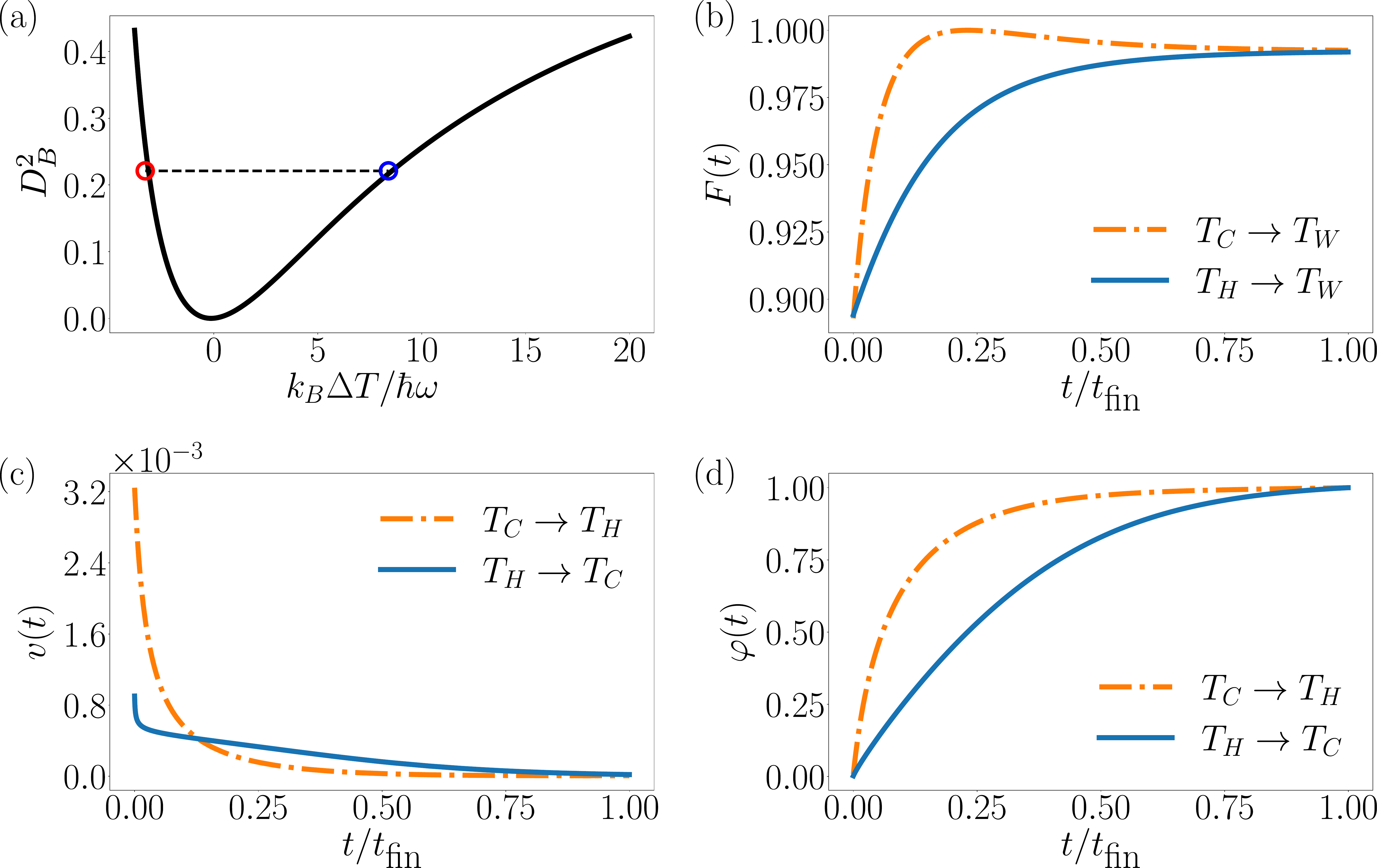}
\caption{Simulation of the protocols for the quantum Brownian particle. (a) Bures distance dependence with the temperature difference with respect to the equilibrium state. (b) Fidelity for the state with respect to the thermal state at $T_W$. The process corresponds to the three-temperature protocol, evolving from $T_C$ to $T_W$ (orange line) and $T_H$ to $T_W$ (blue line). (c) Instantaneous statistical velocity and (d) degree of completion computed for the heating and cooling processes in the two-temperature protocol. In this case, the values of the parameters are $t_{\textrm{fin}} = 5 \cdot 10^{3}, m=1, \Omega = 10^{-3},\Lambda=1, \zeta = 0.1$.}
\label{fig:plots_QBM}
\end{figure*}

\subsection{Quantum Brownian particle in a trap}
The third model introduced to analyze the protocols is a quantum Brownian particle in a harmonic trap, following the results and experiments already performed in the classical case \cite{ibanez2023}. This is the most sophisticated case that is treated in the paper, where all the relevant quantities need to be computed numerically. 

A quantum Brownian particle interacting with a bosonic bath is described by the following Hamiltonian~\cite{pottier2010nonequilibrium,lampo2019quantum,caldeira1983a,caldeira1983b}
\begin{gather}
\begin{aligned}
\label{eqn:ham_QBM}
H 
&=&\frac{p_S^2}{2m_S} + \sum_{i=1}^{n}\frac{\kappa^2_i}{2m_{B_i} \omega_{B_i} }x_S^2 + \phi(x_S) \\
&+& \sum_{i=1}^{n} \hbar \omega_{B_i} a^{\dagger}_{B,i} a_{B,i} - \sum_{i=1}^{n} \kappa_i x_{B,i} x_S,
\end{aligned}
\end{gather}
where the indexes $S,B$ hold for the system and bath operators respectively. Here $m_S$ is the mass of the Brownian particle, $x_S$ its position, $p_S$ its momentum and $\phi(x)$ is a trapping potential. Similarly, $m_{B_i}$, $\omega_{B_i}$, and $x_{B_i}$ are the mass, frequency, and position of the $i$-th bath particle, for all $i=1,\ldots,n$. The factors $\kappa_i$ represent the coupling between the system and the $i$-th bath mode. 

The trapping potential is customarily taken as a harmonic term so that
\begin{equation}\label{eqn:qbm_trap}
    \phi(x) = \dfrac{m}{2}\tilde{\omega}^2x^2,
\end{equation}
for a given trap frequency $\tilde{\omega}$. 

A treatment for the problem can be performed by a Lindblad-equation-like transformation of the equations of motion of the system. The global evolution of the system and bath may be described by a unitary operator, and the state of the system at a given time $t$ is described by
\begin{equation} \label{eqn:rho_QBM}
\rho_S(t)= \textrm{Tr}_B \left\{ U(t) \left(\rho_S(0) {\otimes} \rho_B \right) U^{\dagger}(t) \right\} \equiv e^{\mathcal{L} t}[\rho_S(0)],
\end{equation}
being $\mathcal{L}$ the Liouvillian superoperator of the coherent dynamics. Given the fact that the interaction between the system and the bath is linear and assuming it to be also weak, we can consider a single Lindblad operator $L(T)$ such that \cite{lampo2019quantum}
\begin{eqnarray}\label{eqn:operator_QBM}
L(T)&=&\tilde\alpha(T) x + \tilde\beta(T) p,
\end{eqnarray}
for some parameters $\tilde\alpha(T), \tilde\beta(T) \in \mathbb{C}$. In order to match the coefficients represented in Eq.~\eqref{eqn:operator_QBM} with a general Born-Markov treatment of the problem within the Caldeira-Leggett limit, they must be given by
\begin{equation}\label{eqn:qbm_alpha}
    \tilde\alpha (T)= \dfrac{\left(2m\zeta k_B T\right)^{1/2}}{\hbar} ,
\end{equation}
and
\begin{equation}\label{eqn:qbm_beta}
    \tilde\beta(T) = \dfrac{\zeta}{ \hbar\tilde\alpha}\left(-\dfrac{k_B T}{\hbar \Lambda}+ \dfrac{i}{2 }\right).
\end{equation}
In these relations, $\Lambda$ is the so-called \emph{Lorentz-Drude cutoff} appearing in baths with Ohmic spectral density; $\zeta$ is a damping constant, whose inverse is related to the relaxation scales; $T$ is the temperature of the bath and $m$ the mass of the oscillators. The Caldeira-Leggett limit is satisfied for large temperature and cut-off limits. Under this regime, one recovers the Caldeira-Leggett equation for general diffusion processes in a quantum framework for a quantum Brownian particle \cite{caldeira1983a,caldeira1983b}. Note that the average number of excitations is related to the temperature of the baths via the Bose-Einstein relation as in the previous models.

In Fig.~\ref{fig:plots_QBM} we observe a similar behavior under the two- and three-temperature protocols to the ones performed for both the two-level system and the harmonic oscillator. In this case, the temperature range that we need to consider is even larger, due to the complexity of the bath. All the results are similar, suggesting the general character of our results. One interesting feature is that during the heating up process in the three-temperature protocol [c.f. Fig.~\ref{fig:plots_QBM}(b)] the fidelity reaches a value close to one in a finite time, and then bounces down. This interesting behavior suggests that the system suffers from hysteresis, an interesting feature specially due to the Markovian character of the dynamics.

\subsection{Analysis of the results}\label{subsec:simulations}

In the three-temperature protocol, the fidelity of the state at a given time $t$ has been compared to the thermal state at the intermediate temperature, $T_W$, so that it increases to one, when thermalization takes place. 
As indicated in Sec.~\ref{subsec:three_temps}, the thermal state at the warm temperature, $\rho^{\text{th}}_{\beta_W}$, is chosen to be equidistant to the cold, $\rho^{\text{th}}_{\beta_C}$, and hot states, $\rho^{\text{th}}_{\beta_H}$.
The Bures distance is depicted in Figs. \ref{fig:panel_qubit}(a), \ref{fig:plots_HO}(a), and \ref{fig:plots_QBM}(a) as a function of the temperature difference $\Delta T$ with respect to $T_W$ , meaning that $\Delta T_C = T_C - T_W$ for the initial cold state (heating-up protocol), and $\Delta T_H = T_H - T_W$ for the initial hot state (cooling-down protocol). The temperature of both initial points is represented by a hollow red circle for the case of heating and by a hollow blue circle for the cooling. Both points evolve to the equilibrium thermal state, clearly represented by a minimum. It is worth noticing that the asymmetry in the different protocols is appreciated here. 

The asymmetry in the three-temperature protocols is analyzed by the use of the fidelity between the initial states and the target one, as a function of time.  This is displayed in Figs.  \ref{fig:panel_qubit}(b), \ref{fig:plots_HO}(b), and \ref{fig:plots_QBM}(b). Remarkably, the necessary times to thermalize differ by several orders of magnitude due to the increasing complexity of each configuration. However, the three of them show a similar behavior. This fact is also relevant for the velocity analysis.

Regarding the two-temperature protocol, we shall make use of the instantaneous velocity quantity, Eq. (\ref{eqn:speed}), and the degree of completion, Eq. (\ref{eqn:degree}). Figures \ref{fig:panel_qubit}(c), \ref{fig:plots_HO}(c), and \ref{fig:plots_QBM}(c) represent $v(t)$ for the qubit, the harmonic oscillator, and the Brownian particle, respectively. 
As we anticipated in the previous paragraph, the thermalization times differ by several orders of magnitude. This is also represented by the scale in the velocity axis in the aforementioned plots. \textcolor{black}{The behavior is also different in the three cases (while the heating and cooling velocities cross for the harmonic oscillator and the Brownian particle, for the qubit we  find that the heating velocity is always smaller), which avoids to extract general conclusions from the analysis of $v(t)$. To establish the fastest process one needs to compute the degree of completion}, derived considering the values of the instantaneous velocity, Eqs. (\ref{eqn:length}) and (\ref{eqn:degree}). Figures  \ref{fig:panel_qubit}(d), \ref{fig:plots_HO}(d), and \ref{fig:plots_QBM}(d) show the temporal evolution of the degree of completion and, as it is expected, the functions are similar regardless of the sort of system, and the heating process takes less amount of time to be completed than the cooling. 

Despite their different complexities, in all three cases the thermal kinematics theory finds the same overall behavior in the description of the dynamics towards the equilibrium state. A better intuition of the physical origin of this phenomenon a spectral analysis of the Liouvillian will be presented in Sec.~\ref{sec:spectral}.

\begin{figure*}[t]
\centering
\includegraphics[width=0.8\textwidth]{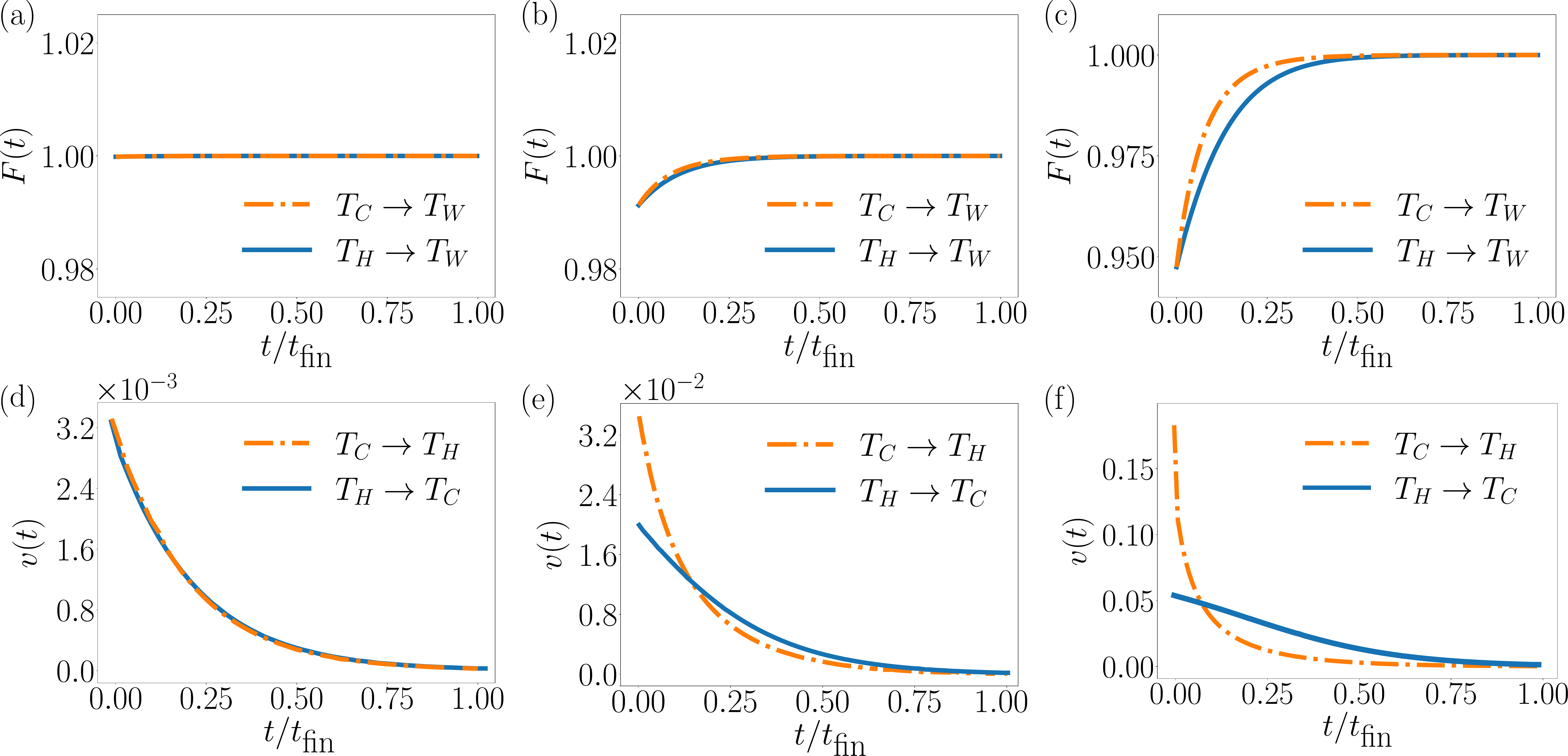}
\caption{Near-temperature simulations for the three-temperature protocol for the harmonic oscillator. (a)--(c) Fidelity with respect to the thermal state at $\bar{n}(\omega, T_W)$ and (d)--(f) instantaneous velocity. In all cases $\omega=1$, and $T_C$ chosen such that $\bar{n}(\omega,T_C) = 1$. The hot temperature is such that $\bar{n}(\omega,T_H) = 1.1$ for (a), (d); $\bar{n}(\omega,T_H) = 2$ for panels (b), (e); and $\bar{n}(\omega,T_H) = 5$ for (c), (f). When $T_C$ and $T_H$ are close, both curves collapse and the processes remain symmetric. However, as the temperature difference increases, heating up and cooling down evolve differently. }
\label{fig:ho_close_temperatures}
\end{figure*}
\subsection{Linear response regime}\label{sec:linear}

We will conclude the section providing a brief comment on the near-equilibrium, i.e., linear regime, for thermal evolution close to the equilibrium temperature in the three-level protocol.

The linear response theory, developed mainly by Kubo~\cite{kubo1957linearresponse,kubo1966}, is the cornerstone to analyze the near-equilibrium behavior in classical thermodynamics. The fluctuation-dissipation theorem states that the fluctuation properties of a system in \textcolor{black}{thermal equilibrium} determine its linear response to an external perturbation~\cite{kubo1966}. In the quantum counterpart, this theorem has been derived for closed quantum systems and recently for open quantum systems~\cite{konopik2019PRR,blair2024PRR}. This extension allows us to apply the existing results from isolated equilibrium systems to open systems, with Lindbladian dynamics \cite{konopik2019PRR,blair2024PRR}. Within this regime, one expects to recover the same as in classical thermodynamics results, where the asymmetry between heating and cooling is absent. That means, for small temperature differences in both protocols, we expect the asymmetry to diminish as the initial deviation is closer to the equilibrium state. 

For the qubit case  this phenomenon is clearly appreciated in the analytical derivation of the fidelity comparing two states, Eq.~\eqref{eqn:QBf0}. For small $\Delta T$, the fidelity is quadratic  
\begin{equation}
\label{eqn:QBF0exp}
F(\rho_{\beta_0}^{\rm th},\rho_\beta^{\rm th})=1-\frac{e^{\beta\hbar\omega}}{8\left(1+e^{\beta\hbar\omega}\right)^2}\left(\frac{\hbar\omega\Delta T}{T^2}\right)^2+{\cal O}\left(\frac{\Delta T}{T}\right)^3,
\end{equation}
i.e., no asymmetry is expected for states close to equilibrium. 

Regarding the simulations for the harmonic oscillator, the results for close temperatures is depicted in Fig. \ref{fig:ho_close_temperatures}(a) and \ref{fig:ho_close_temperatures}(d). As the temperature difference increases, the asymmetry starts to appear, making this discrepancy in both protocols more acute the larger is this gap. Figures \ref{fig:ho_close_temperatures}(b)--(c) and \ref{fig:ho_close_temperatures}(e)--(f) show this behavior. 
Figures \ref{fig:ho_close_temperatures}(a)--(c) represent the fidelity with respect to the thermal state at warm temperature $T_W$, i.e., in the three-temperature protocol. Similarly, Figs.~\ref{fig:ho_close_temperatures}(d)--(f) showcase the asymmetry in the two-temperature scenario, displayed in the velocity needed to reach the opposite state. It is clear that the asymmetry arises as one deviates from equilibrium when the temperature difference increases. 

\begin{figure*}[t]
\centering
\includegraphics[width=0.85\textwidth]{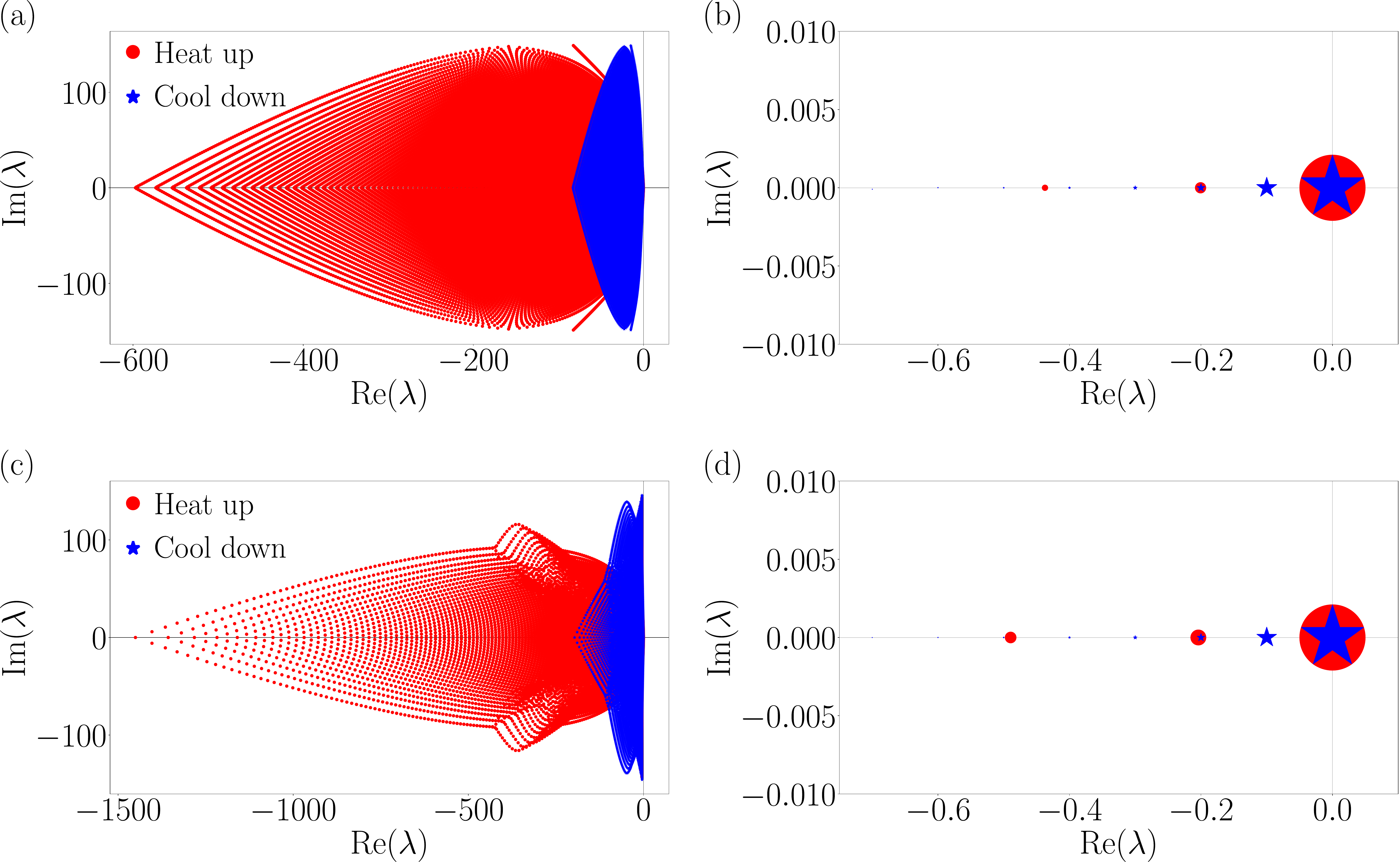}
\caption{Eigenvalues of the Liouvillian operator $\mathcal{L}$ for the harmonic oscillator (a), and a quantum Brownian particle (c) for $\bar{n}(\omega, T_H) = 10$ (red points) and $\bar{n}(\omega, T_C) = 1$ (blue points). 
(b) and (d) represent the first eigenvalues of the respective panels (a) and (c), with a size proportional to the overlap with the thermal state at the opposite temperature,  see Eq. (\ref{eqn:overlap}). In both cases, the truncated dimension of the Hilbert space is $N=150$.}
\label{Fig:Spectrum}
\end{figure*}

\section{Spectral analysis}\label{sec:spectral}
Performing a spectral analysis of the Liouvillian, we can gain intuition about the relaxation time of our models for the heating and cooling protocols, while the dynamics of the system follow Eq. (\ref{eqn:evolution_spectrum}). In Fig. \ref{Fig:Spectrum} the eigenvalues of both the harmonic oscillator and the Brownian particle are displayed. Due to the infinite size of the Hilbert space of the systems, we have used a truncated Fock basis of dimension $N\textcolor{black}{=150}$, large enough to display the general behavior. As discussed in Sec. \ref{sec:framework}, the spectrum of the Lindbladian is composed of eigenvalues whose real part is negative, apart from the null eigenvalue, which determines the stationary state. 
This decomposition does not depend on the initial state we consider but on the parameters defined in the Lindbladian, i.e. the Hamiltonian and jump operators, as well as the constants and variables defined therein, in particular on the temperature of the bath. 
The dependence of the relaxation of an open quantum system on the initial state after the quench is rather given by its overlap with the different decaying modes, $\xi_i$, see Eq.~(\ref{eqn:overlap}).

The spectra for the heating and cooling cases and the different systems are displayed in Fig. \ref{Fig:Spectrum}. For the Lindbladians with higher temperature there is a \emph{spreading} of the eigenvalues towards the negative real axis, see Fig. \ref{Fig:Spectrum}(a) for the harmonic oscillator and Fig. \ref{Fig:Spectrum}(c) for the quantum Brownian particle. This means that, in the heating-up processes, there are many more fast-decaying modes than in the cooling-down counterparts, indicating that heating will be faster. This behavior is in agreement with Fig.~\ref{fig:plots_qubit}(a) for the thermal qubit. In Figs. \ref{Fig:Spectrum}(b) and \ref{Fig:Spectrum}(d), the slowest decay modes, which act as bottlenecks to the dynamics, are plotted with the symbol size being proportional to the overlap with the initial state of the cooling down and heating protocols. The initial state in each case is chosen to be equal to a thermal state at the same temperature as the opposite process. This means that if the cooling down/heating up process is causing the system to evolve to $\bar{n}(\omega, T_C)/\bar{n}(\omega, T_H)$, the initial state will be $\rho^{\text{th}}_{\beta_H}/ \rho^{\text{th}}_{\beta_C}$.

It is clear that in the cooling protocol, there is a higher overlap with slower decay modes in both cases. Moreover, the number of slower modes in the cooling case (represented by the blue stars) is larger than the number of modes in the heating case (red dots), apart from being closer between them and to the null eigenvalue. This spectral analysis provides us with a justification for the asymmetry in all the processes, referring all of them to mere observations of the decaying modes appearing in the spectra of the Liouvillians. 
This explanation allows us to justify and clarify all the results obtained throughout the paper. We recall, however, that this asymmetry in the overlap was not present in the thermal qubit case, which only has a single decaying mode.

\section{Conclusions}\label{sec:conclusions}
In this work, we have investigated an intriguing effect of non-equilibrium open quantum systems: the asymmetry of the time evolution of heating up and cooling down trajectories. By introducing quantum information measures such as the fidelity, the Bures distance, and the quantum Fisher information, we analyzed this phenomenon in two different protocols. The first (three-temperature) protocol involves an intermediate temperature, equidistant between a hotter and a colder one, while the second (two-temperature) protocol works between two absolute temperatures. The measures developed in this work are general and applicable to various other dissipative processes. 

We extended the thermal kinematics to open quantum systems and applied these protocols to three different configurations of increasing complexity: a thermal qubit, a harmonic oscillator coupled to a bosonic heat bath, as well as a canonical model for the quantum Brownian motion. The qubit system provides an analytical description that can be solved exactly for all the studied magnitudes, and allowing for separate interpretations of the two protocols; the other systems are analyzed numerically. 
Our results unequivocally indicate that heating up and cooling down are intrinsically different processes, with heating up always being the fastest for the explored configurations. In the limit of small temperature differences we recover a symmetric behavior in accordance with equilibrium thermodynamics in the quantum regime. 
\textcolor{black}{Note that particular configurations of multilevel systems have been described showing that the asymmetry can be inverted (cooling can be faster than heating)~\cite{VanVu2021}}.

By studying the Liouvillian spectrum of the system, we observe that the eigenvalues spread towards the negative real line as temperature increases. This indicates that for thermal baths at higher temperatures there are more fast-decaying modes, making the evolution faster. Additionally, the overlap between the initial state and the fast-decaying modes 
\textcolor{black}{is larger for the heating up processes.} Despite their simplicity, the proposed configurations can be readily be tested  experimentally is various platforms, e.g., semiconductor qubits~\cite{burkard_semiconductor_2023} or superconducting cavity quantum thermodynamic circuits~\cite{pekola_colloquium_2021}. As systems with higher complexity require longer times to thermalize, harmonic oscillators or quantum Brownian motors are ideal candidates to detect thermalization asymmetries.

\begin{acknowledgments}
The authors would like to thank John Bechhoefer, M. Skotiniotis, G. Menesse, and G. H. Camillo for their insightful comments. The research leading to these results has received funding from
the Ministry for Digital Transformation and of Civil Service of the Spanish Government through the QUANTUM ENIA project call - Quantum Spain project, and by the European Union through the Recovery, Transformation and Resilience Plan - NextGenerationEU within the framework of the Digital Spain 2026 Agenda. Besides we would like to acknowledge the Projects of I+D+i Refs PID2021-128970OA-I00, PID2020-113681GB-I00,  and PID2022-142911NB-I00, financed by MICIN/AEI/10.13039/501100011033 and FEDER “A way to make Europe”, and Projects Ref. A-FQM-175-UGR18, Ref. P20\_00173 and Ref. A-FQM-644-UGR20,  C-EXP-251-UGR23 and through the ``Mar\'{i}a de Maeztu'' Programme for Units of Excellence in R{\&}D CEX2023-001316-M, financed by the Spanish Ministerio de Ciencia, Innovación y Universidades and European Regional Development Fund, Junta de Andalucía-Consejer\'{\i}a de Econom\'{\i}a y Conocimiento 2014-2020.
AT also acknowledges Grant No. FPU20/02835 from Spanish Ministerio de Ciencia, Innovación y Universidades. Finally, we are also grateful for the computing resources and related technical support provided by PROTEUS, the supercomputing center of the Institute Carlos I for Theoretical and Computational Physics in Granada, Spain.
\end{acknowledgments}

\appendix
\section{Classical and Quantum Fisher Information}\label{sec:appendix}
In classical parameter estimation a canonical measure is the classical Fisher information, $I(\theta)$, of a probability density $p(x,\theta)$, defined as
\begin{equation}\label{eqn:classical_fisher_appendix}
    I_{\text{cl}}(\theta) := \int_{-\infty}^{\infty} \left(\dfrac{d \log p(x,\theta)}{d \theta}\right) ^2 p(x,\theta) dx.
\end{equation}
The geometric interpretation of the Fisher information arises from defining a statistical line element, $ds$, such that $ds^2:= I(\theta) d\theta^2$. Therefore, the line element $ds$ can be regarded as a dimensionless distance between probability densities $p(x, \theta)$ and $p(x, \theta + d\theta)$.

Consider a quantum state, $\rho$, parametrized by an $n$-dimension vector $\vec{\theta} = \left(\theta_1,\ldots,\theta_n\right)$, and denoted by $\rho(\vec{\theta})$. For an infinitesimal change in the parameters, one can relate the Bures distance to the \emph{quantum Fisher information matrix} (QFIM), $\mathcal{I}$, so that
\begin{equation}
\left[D_B\left(\rho(\vec{\theta}), \rho (\vec{\theta} + d \vec{\theta})\right)\right]^2
 = \dfrac{1}{4} \sum_{i,j} \mathcal{I}_{ij} dx_i dx_j + \mathcal{O}(dx^4).
\end{equation}
The complete derivation can be found in Ref. \cite{liu2020quantum}. The elements of the QFIM are given by
\begin{equation}\label{eqn:QFIM_general}
    \mathcal{I}_{ij} = \Tr \left[L_{\theta_i} \rho({\vec{\theta}}) L_{\theta_j}\right],
\end{equation}
where $\left\{L_{\theta_i} \right\}_{i=1}^n$ are the \emph{symmetric logarithmic derivative} (SLD) operators for the $k-$th parameter, implicitly defined as
\begin{equation}
    \dfrac{\partial \rho(\vec{\theta})}{\partial {\theta_k}} := \dfrac{L_{\theta_k} \rho(\vec{\theta}) + \rho(\vec{\theta}) L_{\theta_k}}{2}.
\end{equation}
We are only interested in single-parameter estimation, in this case the QFIM reads
\begin{equation}
\mathcal{I}_\theta = \Tr \left[L_\theta^2 \rho (\theta) \right].
\end{equation}
We are intended to obtain an operational expression for the SLD. In the eigenbasis of $\rho(\theta)$, by means of the spectral theorem, the density matrix can be decomposed in terms of its eigenvalues and eigenvectors
\begin{equation}
    \rho(\theta) = \sum_{i=1}^n \lambda_i(\theta) \ket{\lambda_i(\theta)}\bra{\lambda_i(\theta)}.
\end{equation}
Hence, in the eigenbasis of the state $\rho(\theta)$, the SLD operator is simply given by
\begin{equation}
\label{eqn:Ltetha}
L_\theta = 2 \sum_{i,j} \dfrac{\left\langle \lambda_i(\theta) \middle| \dfrac{d\rho (\theta)}{d\theta} \middle| \lambda_j(\theta) \right\rangle}{\lambda_i(\theta) + \lambda_j(\theta)} \left| \lambda_i(\theta) \right\rangle \left\langle \lambda_j(\theta) \right|,
\end{equation}
where $\left\{\ket{\lambda_k(\theta)} \right\}_{k=1}^n$ is the eigenbasis of $\rho (\theta)$ for $\lambda(\theta)_i + \lambda(\theta)_j \neq 0, \ \forall i,j = 1,\ldots,n$.
For our analysis, we use only the time as a parameter, giving 
\begin{equation}
    L_t = 2 \sum_{i,j} \dfrac{\left\langle \lambda_i(t) \middle| \dfrac{d\rho (t)}{dt} \middle| \lambda_j(t) \right\rangle}{\lambda_i(t) + \lambda_j(t)} \left| \lambda_i(t) \right\rangle \left\langle \lambda_j(t) \right|,
\end{equation}
and the QFIM
\begin{equation}
\mathcal{I}_Q = \Tr \left[L_t^2 \rho (t) \right].
\end{equation}

\bibliography{ising1D.bib}

\end{document}